\documentclass[sigconf,natbib,anonymous=false,review=false,screen=true]{acmart}

\usepackage{tcolorbox}
\usepackage{booktabs} % For formal tables
\usepackage{algorithm, algpseudocode}
\usepackage{amsmath}
\usepackage{graphics}
\usepackage{epsfig}
\usepackage{graphicx}
\usepackage{subfigure}
\usepackage{balance}
\usepackage{multirow}
\usepackage{mathrsfs}
\usepackage{acronym}
\usepackage{placeins}
\usepackage[inline]{enumitem}
\usepackage{fancyhdr}
\usepackage{bm}
\usepackage{tabularx}
\usepackage{ragged2e} % for \RaggedRight macro
\usepackage{makecell}
\usepackage{graphicx}
\usepackage{makecell}
\usepackage{colortbl}
\usepackage{xcolor}

\usepackage{xspace}
\AtBeginDocument{%
  \providecommand\BibTeX{{%
    \normalfont B\kern-0.5em{\scshape i\kern-0.25em b}\kern-0.8em\TeX}}}

\newcommand{\ignore}[1]{}

\newcommand{\instrctionsize}{\footnotesize}
\newcommand{\model}{DRE\xspace}
\newcommand{\fullmodel}{\textbf{D}ata-level \allowbreak \textbf{R}ecommendation \allowbreak \textbf{E}xplanation\xspace}
\newcommand{\sbbkgrnd}{\cellcolor{gray!35}}

\setcopyright{none}
\acmConference{}{}{}
\begin{document}

\title{\model: Generating Recommendation Explanations by Aligning Large Language Models at Data-level}
% \title{
% Data-level Recommendation Explanations Based on Large Language Models
% }

\author{Shen Gao}
\authornote{Both authors contributed equally to this research.}
\affiliation{
\institution{University of Electronic Science and Technology of China}
\country{}
% \city{Qingdao}
% \country{China}
}
\email{shengao@pku.edu.cn}

\author{Yifan Wang}
\authornotemark[1]
\affiliation{
\institution{University of Electronic Science and Technology of China}
\country{}
% \city{Qingdao}
% \country{China}
}
\email{yifanwang993w@gmail.com}

\author{Jiabao Fang}
\affiliation{
\institution{Shandong University}
\country{}
% \city{Qingdao}
% \country{China}
}
\email{jiabaofang@mail.sdu.edu.cn}

\author{Lisi Chen}
\affiliation{
\institution{University of Electronic Science and Technology of China}
\country{}
% \city{Beijing}
% \country{China}
}
\email{chenlisi.cs@gmail.com}

\author{Peng Han}
\affiliation{
\institution{University of Electronic Science and Technology of China}
\country{}
% \city{Qingdao}
% \country{China}
}
\email{penghan_study@hotmail.com}

\author{Shuo Shang}
\authornote{Corresponding author.}
\affiliation{
\institution{University of Electronic Science and Technology of China}
\country{}
% \city{Qingdao}
% \country{China}
}
\email{jedi.shang@gmail.com}

\begin{abstract}
Recommendation systems play a crucial role in various domains, suggesting items based on user behavior. 
However, the lack of transparency in presenting recommendations can lead to user confusion. 
In this paper, we introduce \fullmodel (\model), a non-intrusive explanation framework for black-box recommendation models. 
Different from existing methods, \model does not require any intermediary representations of the recommendation model or latent alignment training, mitigating potential performance issues.
We propose a data-level alignment method, leveraging large language models to reason relationships between user data and recommended items. 
Additionally, we address the challenge of enriching the details of the explanation by introducing target-aware user preference distillation, utilizing item reviews. 
Experimental results on benchmark datasets demonstrate the effectiveness of the \model in providing accurate and user-centric explanations, enhancing user engagement with recommended items.
\end{abstract}




\maketitle

\section{Introduction}

Recommendation systems (RecSys) play a pivotal role in learning user preferences and interests by analyzing historical user behavior data~\cite{cheng2016wide,guo2017deepfm,he2017neural,johnson2014logistic}. 
Subsequently, the RecSys recommends relevant items from extensive databases, which are widely used in diverse domains such as e-commerce, news portals, and short video applications~\cite{zhang2021unbert,koren2009matrix,he2016ups,van2013deep}. 
However, the direct presentation of recommended items may inadvertently confuse users, as they may not always comprehend the rationale behind a particular recommendation~\cite{lei2023recexplainer,cheng2022towards,cheng2021learning}. 
This lack of transparency impedes users' inclination to explore the recommended item further~\cite{zhang2020explainable,balog2019transparent,chen2020try}. 
% Consequently, integrating clear and concise recommendation explanations becomes imperative, and it can also increase the likelihood of users making a purchase.
Consequently, interpreting the recommendation results of a black-box recommender model logically has always been an important research direction~\cite{bilgic2005explaining,sharma2013social,tintarev2010designing}.
Most of the existing methods~\cite{xu2023reasons,wang2018reinforcement,zilke2016deepred,peake2018explanation} usually focus on how to employ an additional explanation module to align with the recommendation system, subsequently generating natural language explanations.

\begin{figure}[h]
\centering
  \includegraphics[width=\columnwidth]{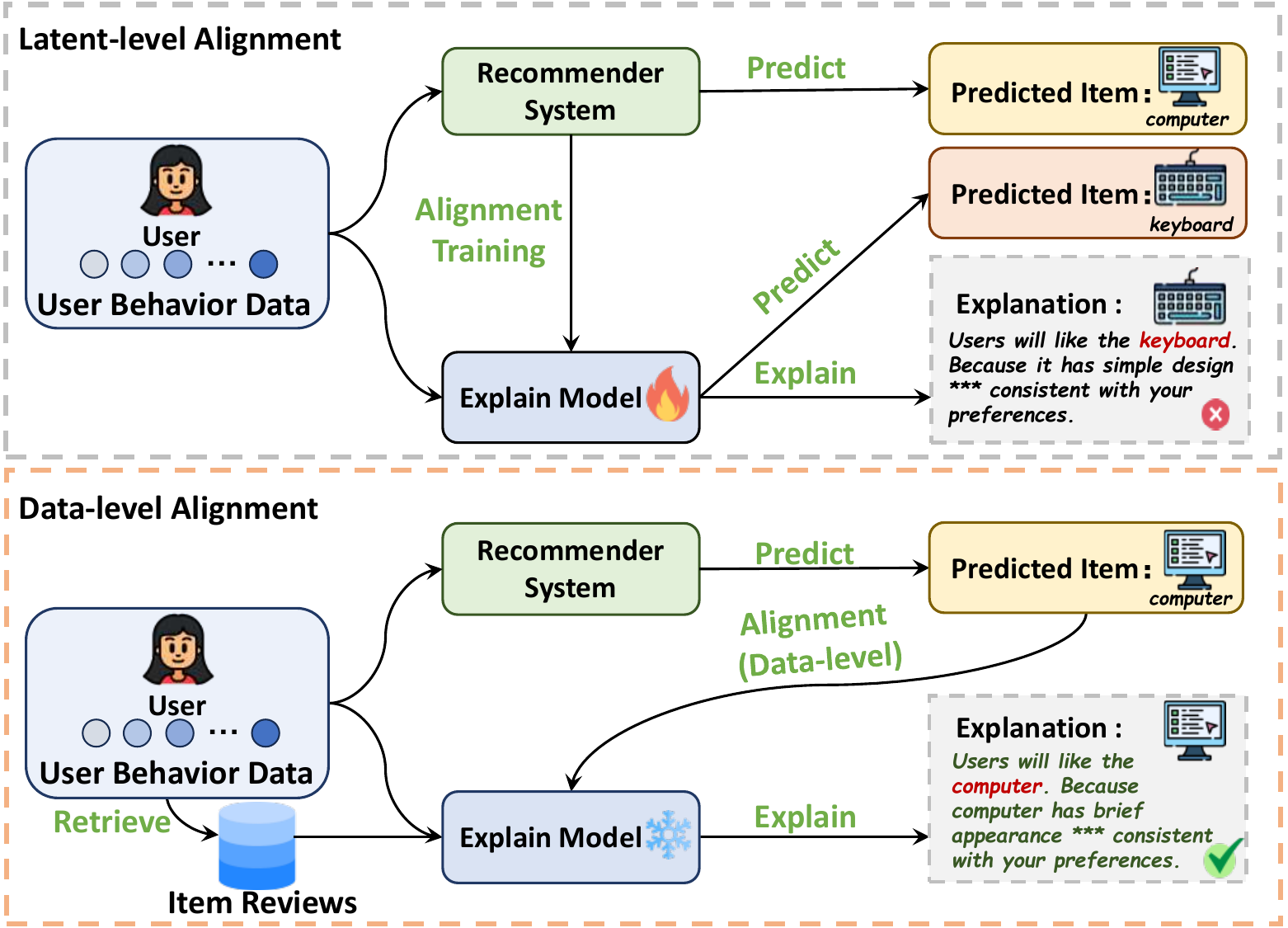}
  \caption{Comparison between existing latent-level and our proposed data-level recommendation explanation method.}
  \label{fig:intro}
\end{figure}

However, there are two key challenges of these methods:
(1) Existing methods~\cite{lei2023recexplainer,xu2024prompting,chen2017attentive,chen2018neural} often involve intrusion into the latent representations within the recommendation model, necessitating modifications to align the explanation and recommendation modules. 
Considering the different training objectives of these two modules, it could adversely affect the performance of both language generation and item recommendation. %(recommendation systems trained for a discriminative task, while language generation models focus on text generation)
Moreover, although these methods aim to align two modules through training, they still cannot guarantee that the recommendation predictions of the two modules are consistent.
In the real-world application, discrepancies between the explained and recommended items may lead to user confusion.
(2) The recommendation system based on ItemID models the co-occurrence relationships among items~\cite{zhang2014explicit,zhang2020distilling,mcauley2013hidden,diao2014jointly,wang2018tem}, lacking an understanding of the specific semantic information about the items, such as the specific purposes of the products or the particular scenarios in which users use them.
However, simply aligning the two modules cannot provide the explanation module with rich semantic information. 

In this paper, we propose the \fullmodel (\model) which can be applied to any black-box recommendation model without accessing intermediate representations or modifying the model.
To avoid modifying the recommendation system, we propose a \textit{data-level alignment method} to align the explanation module and the recommendation model.
Figure~\ref{fig:intro} shows the comparison between our proposed paradigm and existing methods.
Since the large language models (LLMs) have shown strong reasoning capability in many tasks~\cite{wei2022emergent,mann2020language,dong2019unified,devlin2018bert,radford2018improving,zhao2023survey,xi2023rise}, we propose to employ the LLM to reason the relationships between the user's historical data and recommended items.
Specifically, we feed the input user historical behavior data used by the recommendation model and the recommended item to the LLM.
And we leverage the internal knowledge of LLM to find a reasonable relationship between the user preference and the attributes of the recommended item.
This data-level alignment method can align these two modules without requiring any internal representation or intermediate result of the recommendation model, and it can easily be plugged into any RecSys.

For the second challenge, due to the limited detailed information of item descriptions, relying solely on item descriptions for inferring relationships between items can sometimes be challenging in uncovering implicit relationship information. 
Therefore, we propose utilizing the reviews of the items purchased by users and the reviews of the target recommended items to enhance the explanation module's understanding of user preferences and the semantics of target items. 
Since there is a lengthy of reviews for items that users have purchased, extracting relevant information from these reviews and generating explanations that better align with user preferences is a challenge. 
Thus, we introduce the \textit{target-aware user preference distillation} method, which leverages the understanding and reasoning capabilities of LLM, employing semantic matching to extract target-aware information from reviews on items previously purchased by users. 
Finally, by incorporating the extracted target-aware information, we generate explanations for the recommended target items.
Experiments conducted on several benchmark datasets from recommendation systems demonstrate that our proposed \model generates explanations accurately describing aspects that users care about, thereby enhancing user interest in recommended items.

Our contributions are as follows:
\begin{enumerate*}[label=(\roman*)]
\item We propose \model, an LLM-based non-intrusive explanation framework for recommendation systems.

\item We propose a data-level alignment method to align the explanation module and the recommendation model.

\item We introduce a target-aware user preference distillation method to distill user-related information from item reviews.

\item Experimental results on benchmark datasets illustrate the advantage of \model in terms of the accuracy of explanation.
\end{enumerate*}

\section{Related Work}

% \textbf{Recommendation Explanation.}
Explaining the black box of recommender systems has long been a prominent research direction in the field of recommender systems.
% Accurate and satisfactory explanation can improve users' trust and build bridges between users and recommender systems.
Current research can be mainly divided into two categories. 
The first category focuses on identifying the most critical factors influencing recommendation results\cite{chen2016learning,pan2020explainable}. 
\citet{tan2021counterfactual} formulate an optimization problem to generate minimal changes to item aspects, thereby altering the recommended result. 
These aspects can be viewed as the composition of an explanation detailing why the original item is recommended. 
% \citet{gao2019explainable} propose an explainable deep model and optimizes key variables based on accuracy to capture user’s interest.
\citet{zilke2016deepred,lakkaraju2017interpretable,shrikumar2017learning} define information-based measures to identify the attributes that the model utilizes from the input to generate explanations.
The second category mainly focuses on training a surrogate model to explain the target model. 
For example, \citet{wang2018reinforcement} propose a reinforcement learning framework that gets rewards from the environment and modifies recommendation explanation.
% \citet{datta2016algorithmic,gao2019explainable} propose frameworks that generate compact decision sets that explain the target items.
% \cite{sharma2013social} presents a probabilistic model to generate an explanation.
\citet{ma2019jointly,catherine2017explainable} propose a framework for generating explanations based on the knowledge graph. 
\citet{lei2023recexplainer} employ LLMs as surrogate models, aiming to mimic and understand target recommender models by leveraging both natural language and latent spaces. After alignment, LLMs can generate target items and provide recommendation explanations.
However, existing methods either rely solely on a few entity words or keywords as explanations or employ complex fine-tuning approaches to generate natural language explanations. 
It makes the explanations not natural or complex to use, which requires fine-tuning or modification of existing recommendation systems.

\section{\model Methodology}

% \subsection{Overview}\label{sec:overview}

\begin{figure}[bht]
\centering
  \includegraphics[width=\columnwidth]{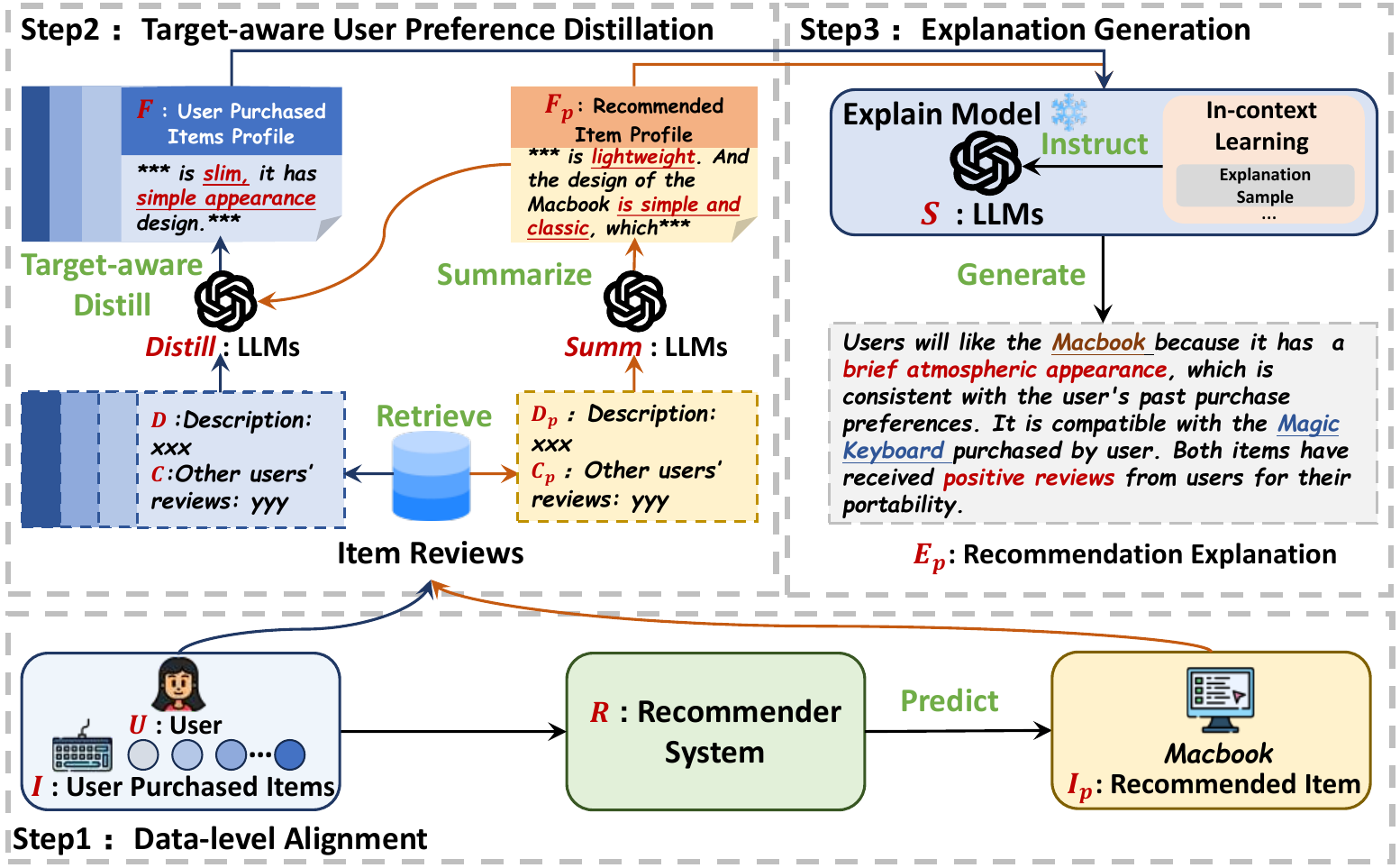}
  \caption{Overview of \model, which firstly align the explanation module and recommender with \textbf{Data-level Alignment}, and then generate the explanation by incorporating details of target from \textbf{Target-aware User Preference Distillation}}
  \label{fig:model}
\end{figure}

In this section, we detail the \fullmodel (\model).
An overview of \model is shown in Figure~\ref{fig:model}.
% \model has three main parts:
% (1) \textbf{Data-level Alignment.} We first align the behavior of the explanation module and recommendation module at the data level, ensuring the explanation module can generate natural language explanations consistent with the recommended results. 
% (2) \textbf{Target-aware User Preference Distillation.} To make the explanation align with user preferences, we extract target-aware information from reviews on items previously purchased by users. 
% (3) \textbf{Explanation Generation.} We generate logically reasonable recommendation explanations by integrating the semantic information that users care about.

\subsection{Data-level Alignment}\label{sec:align}

In order to generate precise explanations for recommended results, we propose a data-level alignment method to achieve behavioral consistency between the recommendation module and the explanation module. 
Given a list of items $I = \{I_1, I_2, \dots, I_N\}$ which is purchased by the user $U$, the recommendation model $R$ predicts items $I_p$ that the user $U$ might find interesting. 
To achieve alignment between the recommendation module and the explanation module, previous methods typically fine-tune the explanation module to perform the recommendation prediction task as well, generating items $I_p$ consistent with the predictions of the recommendation model $R$. 
However, this approach inevitably reduces the text generation capability of the explanation module due to changes in its model structure and parameters.
In this paper, we propose leveraging the in-context learning and reasoning abilities of LLM to align the explanation module with the recommendation module. 
Given inputs $I$ and outputs $I_p$ that are consistent with the recommendation model $R$, LLM can learn this prediction pattern in the context and explore the associated relationships to generate natural language explanations.

\subsection{Target-aware User Preference Distillation}\label{sec:distill}

Relying solely on item IDs and item descriptions for recommendation explanations may fail to capture the details or user actual experiences of the item, which are crucial for users. 
Therefore, we propose to incorporate the reviews of user-purchased items $I$ and the target item $I_p$ predicted by the recommendation model $R$ to assist the explanation model in obtaining more item detail information. 
Given a purchased item $I_i$ of user $U$, we retrieve $M$ reviews $C^i = \{C^i_1, C^i_2, \dots, C^i_M\}$ of item $I_i$ written by \textit{other} users from the database, where each $C^i_1$ represents a paragraph of natural language product review.
Then, we can retrieve $M$ user reviews for each purchased item $I_i$ of user $U$, and then obtain a review set $C = \{C^1, C^2, \dots, C^N\}$ which contains $M \times N$ reviews of other users.
Similarly, we can also retrieve $M$ reviews for the target item $I_p$ denoted as $C^p= \{C^p_1, C^p_2, \dots, C^p_M\}$ which is also written by other users.
In this paper, we assume that the item characteristics described in the review set $C$ are the key features that user $U$ cares about, since the user $U$ has bought these items.
Therefore, we need to perform semantic matching between $C$ and $C^p$ to extract those item features that are both of interest to the user in the past purchased items and possessed by the target product $I_p$.
We propose the \textit{target-aware user preference distillation} method, which involves matching the target item reviews $C^p$ with $C$ to extract valuable information for generating recommendation explanations.

Since the description and reviews of items are usually quite long, and not all the information is helpful for generating recommendation explanations.
For the target item $I_p$, we first construct an overview item profile $F_p$ to distill the useful item features. 
We use the product description $D_p$ and reviews information $C^p= \{C^p_1, C^p_2, \dots, C^p_M\}$ of $I_p$ as input and prompt the LLM to generate an item profile $F_p$:
\begin{equation}\label{eq:summ}
    F_p = \text{Summ}\left(\{C^p_1, C^p_2, \dots, C^p_M\}, D_p\right),
\end{equation}
where $F_p$ contains both the basic information of the target item and user usage experiences and $\text{Summ}$ is an LLM-based module that is prompted by the following instructions:
\begin{tcolorbox}[colback=black!1!white,colframe=black!57!white,boxsep=1pt,left=1pt,right=1pt,top=1pt,bottom=1pt]
\instrctionsize
You are given item's description and reviews. 
Response item profile using the following format:
\\item: \{item name\} 
\\description: \{item description\}
\\other users' reviews: \{item reviews\}
\\Extract key features from reviews.
\end{tcolorbox}

However, not all the product features mentioned in $F_p$ may be of concern to the user $U$. 
Therefore, we need to extract product features that user $U$ care about from $C = \{C^1, C^2, \dots, C^N\}$ associated with user behavior. 
Specifically, we use the item profile $F_p$ of the target item to filter reviews in set $C^i$ of item $I_i$:
\begin{equation}\label{eq:distill}
    F_i = \text{Distill}\left(F_p, \{C^i_1, C^i_2, \dots, C^i_M\}, D_i\right),
\end{equation}
where $D_i$ is the item description of item $I_i$, and $\text{Distill}$ is an LLM-based module that is prompted by the following instructions:
\begin{tcolorbox}[colback=black!1!white,colframe=black!57!white,boxsep=1pt,left=1pt,right=1pt,top=1pt,bottom=1pt]
\instrctionsize
Finish history item profile using relevant features with recommended item, strictly adhere to the following format when responding:
\\history item: \{item name\}
\\genre: \{item genre\}
\\relevant information: \{item information\}
\\other users' reviews: \{reviews\}
\\which relevant information mainly describes similarities between history item and recommended item, and summarize other users' reviews;
\end{tcolorbox}
By integrating these two parts of information, we obtain the target-aware item profiles $F = \{F_1, F_2, \dots, F_N\}$ for the items the user $U$ has purchased.

\subsection{Explanation Generation}\label{sec:gen}

Finally, we integrate the item profile $F_p$ of the target item with the item profiles $F = \{F_1, F_2, \dots, F_N\}$ of the items the user has purchased. 
We employ an in-context learning approach and instruct the LLM as follows to generate a logically coherent recommendation explanation that aligns with the recommendation system $R$ and corresponds to user attention preferences:
\begin{equation}
    E_p = S\left(F_p, \{F_1, F_2, \dots, F_N\}\right),
\end{equation}
where $S$ is an LLM-based module to generate the recommendation explanation which is instructed by the following instructions:
\begin{tcolorbox}[colback=black!1!white,colframe=black!57!white,boxsep=1pt,left=1pt,right=1pt,top=1pt,bottom=1pt]
\instrctionsize
Now you are a recommendation assistant, combined with history relevant items, write an explanation of the recommended item. The format of response is as below:
\\item: \{recommended item\}
\\recommend reason: \{reason\}
\end{tcolorbox}
\section{Experimental Setup}

% \subsection{Implementation Details}

% In our experiments, all \model-C variants and the ChatGPT baseline use the gpt-3.5-turbo-0613 version, and the \model-M variant and \texttt{Mistral} baseline use the Mistral $8 \times 7B$ version which is open-sourced.
% We use the temperature \todo{1.0} in our experiments.
% And we update the memory modules of agents in \model after each turn, meaning that only the suggestions and experiences from the previous turn are retained.

\subsection{Evaluation Metrics \& Dataset}

% To quantitatively measure the performance of \model, we propose two evaluation metrics in our paper:
We employ two evaluation metrics in our experiments:
(1) \textbf{Aspect Score}:
We assume that the aspects mentioned in the review $C^p_U$ of the target item $I_p$ written by user $U$ are crucial to the user.
We use the review $C^p_U$ as a reference of the explanation $E_p$.
We first employ the LLM to extract aspects of the review $C^p_U$.
%and the recommendation explanations $E_p$
Subsequently, we measure the alignment between recommendation explanations $E_p$ and user preferences by calculating the extent of the aspect overlap between $E_p$ and $C^p_U$: $\text{Aspect\_Score} = \textstyle\frac{1}{N_{a}}{\sum_{i=1}^{N_{a}}hit(i)} \in [0,1],$
where $N_a$ is the number of aspects in the user review $C^p_U$. 
And when the aspect $i$ in the explanation is semantically the same as the aspect in the recommendation explanations $E_p$ then $hit(i)=1$, otherwise, $hit(i) = 0$.
(2) \textbf{Rating Score}: 
Following \citep{lei2023recexplainer}, to directly evaluate the quality of the generated explanation, we implement a three-level scoring criteria to
quantitatively evaluate the response from the LLM:
\begin{enumerate*}[label=(\roman*)]
    \item RATING-1: Poor Explanation, using chunks of original sentence from provided data.
    \item RATING-2: Acceptable Explanation, consider only one aspect of user history and reviews, explaining unrelated items together.
    \item RATING-3: Satisfactory Explanation.
\end{enumerate*}
We employ the LLM to evaluate the generated explanation following these criteria and calculate the average rating score over all the testset.
% \begin{equation}
% \text{Rating\_Score} = \frac{\sum_{i=1}^{N_{r}}N_{i}*Rating_i}{N},
% \end{equation}
% where $N$ is the number of all samples, $N_i$ represents the quantity of RATING-i, $N_r$ represents the count of RATING categories.
% where $N$ is the number of all samples, $N_i$ is the number of $Rating_i$, $N_r$ is the rating score. 

% \subsection{Dataset}

We employ several categories of the Amazon Review dataset~\cite{Ni2019JustifyingRU}, including Cell Phones \& Accessories, Clothing Shoes \& Jewelry, and Home \& Kitchen.
% Intuitively, in order to better capture user preferences, we model user preferences only using positive user reviews.
% Cell Phones \& Accessories contains 12,467 users, 6,977 items and 38,729 reviews.
% Home \& Kitchen contains 16,102 users, 1,590 items and 20,277 reviews.
% Clothing Shoes \& Jewelry contains 19,310 users, 3,746 items and 24,712 reviews.
% To construct the user purchase history, we limit the items sequence to a minimum of 4 items on Clothing Shoes \& Jewelry, Home \& Kitchen, and a minimum of 3 items on Cell Phones \& Accessories. 
% The last item is then used as the prediction target item.
We select 100 samples in each category as testset and ensured that the length of their browsing history is not less than 2 and each item has associated reviews.
% To evaluate our proposed method, we randomly select 100 samples on each category to construct a test set.
% We filtered the data by removing the sample of items with fewer than 2 user-purchased items and no accompanying reviews from users.

\subsection{Comparison Methods}

We compare \model to a state-of-the-art LLM-based recommendation explanation method and several LLMs:
\begin{enumerate*}[label=(\roman*)]
\item \textbf{\texttt{RecExplainer}}~\cite{lei2023recexplainer} introduces an explanation approach by leveraging LLM, which employs three methods - behavior alignment, intention alignment, and hybrid alignment - in the latent spaces.

\item \textbf{\texttt{ChatGPT}}~\footnote{\url{https://chat.openai.com/}} is a closed-source LLM from OpenAI. We use the version gpt-3.5-turbo-0613. 
We conduct explanation as a prompt learning method that uses a single instruction with the same input data as our \model.

\item \textbf{\texttt{Mistral}}~\cite{Mixtral87B} is an open-source LLM and we use the mixture-of-experts version with 8 $\times$ 70 billion parameters, and use the same prompt as \texttt{ChatGPT}.
% that has been trained using both supervised fine-tuning and reinforcement learning with human feedback~\cite{touvron2023llama}.
\end{enumerate*}

We employ two variants of \model: \textbf{\model-C} and \textbf{\model-M} with \texttt{ChatGPT} and \texttt{Mistral} as the backbone respectively. 
And we also employ several ablation models:
\begin{enumerate*}[label=(\roman*)]
\item \textbf{\texttt{\model w/o Rev.}}: We remove all the reviews in our model and only use the description as input.

\item \textbf{\texttt{\model w/o Dist.}}: We directly summarize the description and reviews for the user-purchased item using Equation~\ref{eq:summ} without using the $\text{Distill}$ method in Equation~\ref{eq:distill}.

\item \textbf{\texttt{\model w/o Dist.+$F_p$}}: Based on \texttt{\model w/o Dist.}, we also directly utilize the description and reviews of the target item without using the $\text{Summ}$ method in Equation~\ref{eq:summ}.

\item \textbf{\texttt{\model w/ $F_p$}}: We directly generate the explanation by using the $F_p$ as input to LLM, without using any information from user-purchased items.
\end{enumerate*}
% All the ablation studies are conducted on \textbf{\model-C}.

\begin{table}[t]
\centering
% \scriptsize
\caption{Recommendation explanation performance comparison. $\ddagger$ indicates significant improvement over \texttt{ChatGPT} with $p \le 0.01$ according to a Student's t test.}
\label{table:main_result}
\resizebox{1.0\columnwidth}{!}{
\begin{tabular}{lcc|cc|cc}
\toprule
\multirow{2}{*}{Method} & \multicolumn{2}{c|}{Home \& Kitchen}  &  \multicolumn{2}{c|}{Clothing Shoes \& Jewelry} & \multicolumn{2}{c}{Cell Phones \& Accessories}   \\ 
 & Asp ($\uparrow$) & Rat ($\uparrow$) & Asp ($\uparrow$) & Rat ($\uparrow$) & Asp ($\uparrow$) & Rat ($\uparrow$) \\
\midrule
\texttt{RecExplainer}
& 0.6057 & 2.64 & 0.5628 & 2.68 & 0.6028 & 2.64
\\

\texttt{Mistral}
& 0.7028 & 2.65 & 0.5757 & 2.79 & 0.6571 & 2.00 
\\

\texttt{ChatGPT}
& 0.6971 & 2.51 & 0.6362 & 2.86 & 0.6229 & 2.67
\\

\midrule

\model-M
& 0.7142 & 2.68 & 0.6485 & 2.89 & 0.6857 & 2.57
\\

\model-C
& \textbf{0.7714}$^{\ddagger}$ & \textbf{2.88}$^{\dagger}$ & \textbf{0.6728}$^{\ddagger}$ & \textbf{2.94}$^{\ddagger}$ & \textbf{0.7400}$^{\ddagger}$ & \textbf{2.90}$^{\ddagger}$
\\
\sbbkgrnd \texttt{\model-C w/o Rev.}  & \sbbkgrnd 0.6914 & \sbbkgrnd 2.64 & \sbbkgrnd 0.6400 & \sbbkgrnd 2.65 & \sbbkgrnd 0.6542 & \sbbkgrnd 2.66 \\
\sbbkgrnd \texttt{\model-C w/o Dist.}    & \sbbkgrnd 0.7100 & \sbbkgrnd 2.78 & \sbbkgrnd 0.5971 & \sbbkgrnd 2.72 & \sbbkgrnd 0.7142 & \sbbkgrnd 2.87 \\
\sbbkgrnd \texttt{\model-C w/o Dist.+$F_p$} & \sbbkgrnd 0.7371 &\sbbkgrnd 2.75  & \sbbkgrnd 0.6657 & \sbbkgrnd 2.83 & \sbbkgrnd 0.7200 & \sbbkgrnd 2.87 \\ 
\sbbkgrnd \texttt{\model-C w/ $F_p$} & \sbbkgrnd 0.7385 &\sbbkgrnd 1.64  & \sbbkgrnd 0.5814 & \sbbkgrnd 2.06 & \sbbkgrnd 0.6585 & \sbbkgrnd 2.03 \\ 
\bottomrule
\end{tabular}
}
\end{table}

\section{Experimental results}

\subsection{Main Results}

Table~\ref{table:main_result} shows the performance of our proposed \model and baselines in terms of two metrics.
We can find that \model shows superior performance in terms of all metrics compared to the SOTA method \texttt{RecExplainer} and LLM backbone respectively. 
This phenomenon indicates that compared to the latent-level alignment, our data-level alignment is capable of generating explanations of higher quality. 
Furthermore, the proposed target-aware user preference distillation method can assist the model in capturing more user preference information.
% This is because LLMs lack reviews and can only reveal a limited relationship between user-purchased items and target item based solely on descriptions.
% This suggests that with the help of our proposed \model framework, distilling reviews and descriptions of user-purchased items can help LLMs extract more target-aware information that aligns with user preferences.

\subsection{Ablation Study} 

To evaluate the effectiveness of each module in \model, we also conduct ablation studies with model \model-C, and the results are shown in Table~\ref{table:main_result}.
We found that the \texttt{\model w/o Rev.} method achieved lower scores compared to other ablation models, indicating the effectiveness of integrating review information in our approach.
Additionally, due to the complexity of information in reviews, generating meaningful explanations requires extracting target-aware information. 
Therefore, \texttt{\model w/o Dist.} also exhibited lower performance after removal Distill module from \model.

\subsection{Case Study}

\begin{figure}[bht]
\centering
  \includegraphics[width=\columnwidth]{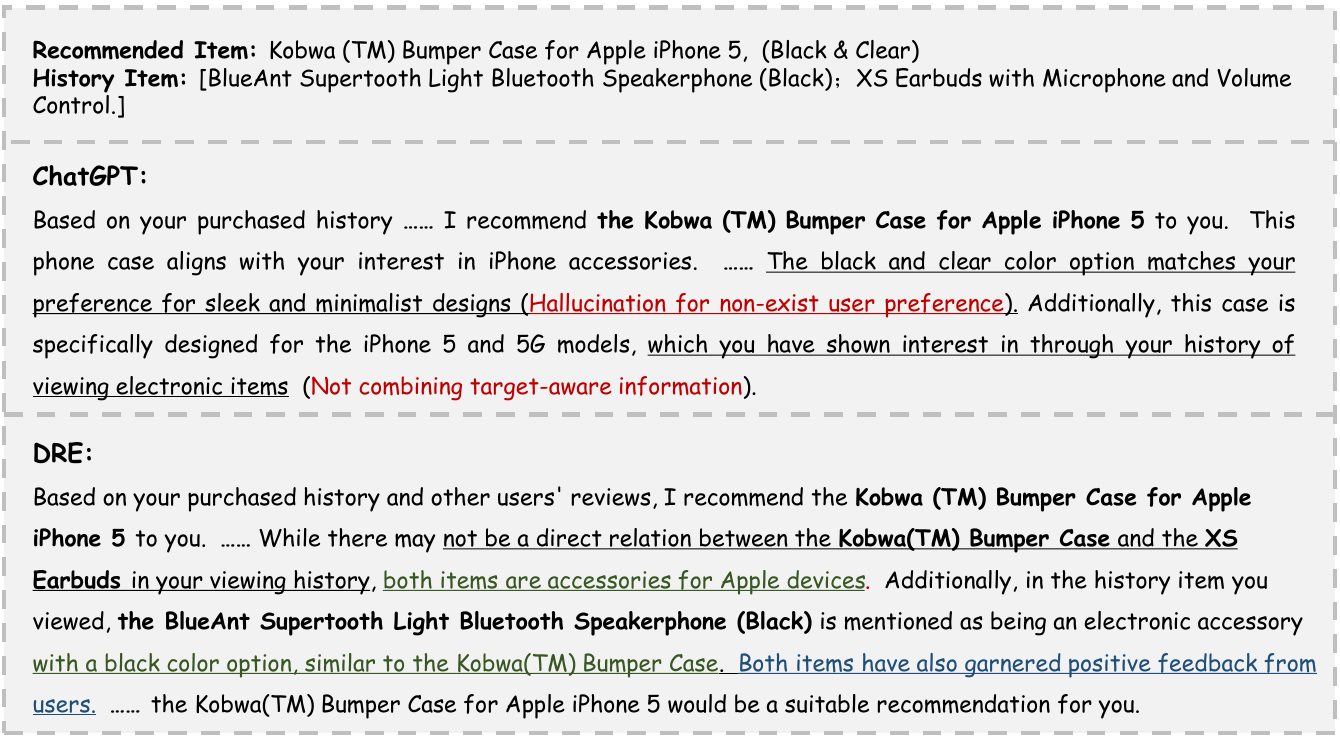}
  \caption{An example of explanation by \model and \texttt{ChatGPT}.}
  \label{fig:good_case}
\end{figure}

Figure~\ref{fig:good_case} shows an example of recommendation explanations generated by \texttt{ChatGPT} and \model based on information about user-purchased items and recommended item.
The bold text in the explanation indicates the recommended item and user-purchased items.
The text in red indicates the shortcomings of the explanation.
The text in green shows target-aware information. 
The text in blue represents consistent reviews from user $U$ for user-purchased items and recommended item.
% From this case, we can find that ChatGPT fails to extract target-aware information and lacks persuasive information to demonstrate that the explanation align with the user's preferences.
From this case, we can find that \texttt{ChatGPT} fails to establish convincing and reasonable relationships between recommended item and user preferences.
And \model provides target-aware information that is persuasive and aligns with user preferences.
This observation demonstrates that our proposed target-aware user preference distillation can effectively filter target-aware information from reviews and descriptions.

\section{Conclusion}

In this paper, we introduced \fullmodel (\model), a non-intrusive explanation framework for black-box recommendation models. 
Our data-level alignment method, leveraging large language models, seamlessly integrates with any recommendation system without intrusive modifications. 
Additionally, our proposed target-aware user preference distillation method enhances semantic understanding using item reviews. 
Experimental results demonstrate effectiveness of \model in providing accurate and user-centric explanations, contributing to the improvement of recommendation system interpretability and user engagement.

% \section*{Reproducibility}
% To facilitate the reproducibility of the results reported in this paper, the
% code and data used are available at \url{https://anonymous.4open.science/r/xxxxxxxxxxxxx}.

% \clearpage
\bibliographystyle{ACM-Reference-Format}
\bibliography{reference}

\end{document}